\documentclass[sigconf, nonacm]{acmart}

\copyrightyear{2021}
\acmYear{2021}
\setcopyright{iw3c2w3}
\acmConference[WWW '21 Companion]{Companion Proceedings of the Web Conference 2021}{April 19--23, 2021}{Ljubljana, Slovenia}
\acmBooktitle{Companion Proceedings of the Web Conference 2021 (WWW '21 Companion), April 19--23, 2021, Ljubljana, Slovenia}
\acmPrice{}
\acmDOI{10.1145/3442442.3451369}
\acmISBN{978-1-4503-8313-4/21/04}

\begin{document}

\title{BIP! DB: A Dataset of Impact Measures for Scientific Publications}

\author{Thanasis Vergoulis}
\email{vergoulis@athenarc.gr}
\affiliation{%
  \institution{IMSI, ATHENA RC}
  \city{Athens}
  \country{Greece}
}

\author{Ilias Kanellos}
\email{ilias.kanellos@athenarc.gr}
\affiliation{%
  \institution{IMSI, ATHENA RC}
  \city{Athens}   
  \country{Greece}
}

\author{Claudio Atzori}
\email{claudio.atzori@isti.cnr.it}
\affiliation{%
  \institution{ISTI, CNR}
  \city{Pisa}
  \country{Italy}
}

\author{Andrea Mannocci}
\email{andrea.mannocci@isti.cnr.it}
\affiliation{%
 \institution{ISTI, CNR}
  \city{Pisa}
  \country{Italy}
}

\author{Serafeim Chatzopoulos}
\email{schatz@athenarc.gr}
\affiliation{%
 \institution{IMSI, ATHENA RC}
 \city{Athens}   
 \country{Greece}
}

\author{Sandro La Bruzzo}
\email{sandro.labruzzo@isti.cnr.it} 
\affiliation{%
  \institution{ISTI, CNR}
  \city{Pisa}
  \country{Italy}
}

\author{Natalia Manola}
\email{natalia@di.uoa.gr}
\affiliation{%
  \institution{OpenAIRE}
  \city{Athens}   
  \country{Greece}
}

\author{Paolo Manghi}
\email{paolo.manghi@isti.cnr.it}
\affiliation{%
  \institution{ISTI, CNR}
  \city{Pisa}
  \country{Italy}
}

\renewcommand{\shortauthors}{Vergoulis, et al.}

\begin{abstract}
  The growth rate of the number of scientific publications is constantly increasing, creating important challenges in the identification of valuable research and in various scholarly data management applications, in general. 
  In this context, measures which can effectively quantify the scientific impact could be invaluable. 
  In this work, we present BIP! DB, an open dataset that contains a variety of impact measures calculated for a large collection of more than $100$ million scientific publications from various disciplines.
\end{abstract}


\maketitle

\section{Introduction}
\label{sec:intro}

The growth rate of the number of published scientific articles is constantly increasing~\cite{growth2}.  
At the same time, studies suggest that, among the vast number of published works, many are of low impact or may even contain research of questionable quality~\cite{ioannidis2005most}. Consequently, identifying the most valuable publications for any given research topic has become extremely tedious and time consuming. 

Quantifying the impact of scientific publications could facilitate this and other related tasks, which make up the daily routine of researchers and other professionals of the broader scientific and academic community. 
For instance, most contemporary search engines for research publications (e.g., Google Scholar, Semantic Scholar) combine keyword-based relevance with a scientific impact measure (usually citation counts) to rank their search results, in an attempt to help their users prioritise reading for literature review. 

However, many impact measures, which are widely used
in various applications, have inherent drawbacks. 
For instance, citation counts cannot differentiate 
citations based on the importance of the citing articles. 
This is an important drawback since citation counts may, 
for example, present a publication in 
a predatory journal, heavily cited by other
trivial works, as invaluable, while disregarding
the seminal importance of an otherwise
sparsely cited publication that has influenced (and is 
cited by) a breakthrough article. For the same reason, citation counts are also vulnerable to various malpractices (e.g., excessive self-citation, citation cartels).

Another important issue, often overlooked by most existing academic search engines, is the importance of capturing publication impact from a broader set of perspectives.
It is an oversimplification to rely on 
one impact measure only, 
since there are many different aspects of scientific impact~\cite{bollen2009principal,kanellos2019impact}.
Indicatively, a publication's \emph{influence} is its overall, long-term importance, which can be calculated based on its whole citation history; its \emph{popularity}, on the other hand, is the publication's current attention in the research community, which is indicative of its expected short-term impact.
These impact aspects are not entirely correlated and each one may be preferable for different applications. 
Consider, for example, an experienced researcher who needs to reexamine a topic of interest to learn about its latest developments. Ranking articles based on their popularity would be preferable for her.
On the other hand, a young researcher wanting to delve into the same topic to prepare a survey would prefer to rank the relevant articles based on their influence.
Although using citation counts would satisfy the needs of the young researcher to an extent, they would fail to help the needs of the experienced one, since citation counts are biased against recent articles. 
This is because any recent article (irrespective of its current attention in the research community) usually requires months or even years to receive its first citations~\cite{smith200930}, and eventually gain momentum.

Overlooking the above aspects is problematic. 
On the one hand, recently published research may be at the center of the current attention of the scientific community, although this is not reflected by the traditional measures. 
On the other  hand, scientific impact should not be examined through a limited set of measures. 
This is important not only because a larger set of measures captures a wider range of impact aspects, providing a more complete picture about a publication's impact; since any individual measure used for research assessment is bound to be abused, based on Goodhart's law\footnote{Also known as Campbell's law (an example of ``Cobra effect'').}, consulting multiple measures can work as a countermeasure to reduce the effects of relevant attacks/malpractices. 
Taking all these into consideration, it is part of the good practices in research assessment to consider a range of quantitative measures as inclusive as possible (e.g., this is
also emphasized in the Declaration on Research Assessment\footnote{The Declaration on Research Assessment (DORA), \url{https://sfdora.org/read}}). 

In this work, we present BIP! DB, an open dataset that contains various impact measures calculated for more than $104M$ scientific articles, taking into consideration more than $1.25B$ citation links between them. The production of this dataset is based on the integration of three major datasets, which provide citation data: OpenCitation's~\cite{opencitations} COCI dataset, Microsoft Academic Graph, and Crossref. Based on these data, we perform citation network analysis to produce five useful impact measures, which capture three distinct aspects of scientific impact. This set of measures makes up a valuable resource that could be leveraged by various applications in the field of scholarly data management. 

The remainder of this manuscript is structured as follows. 
In Section~\ref{sec:bgr} we discuss various aspects of
scientific impact and present the impact measures in the 
BIP! DB dataset. In Section~\ref{sec:workflow} we provide
technical details on the production of our dataset and in
Section~\ref{sec:disc} we empirically show 
how distinct the different measures are and elaborate on 
some interesting issues about the dataset's potential uses and extensions. Finally, in Section~\ref{sec:concl} we conclude the work.    

\section{Impact Aspects \& Measures}
\label{sec:bgr}

As mentioned in Section~\ref{sec:intro}, there are
many perspectives from which one can study scientific
impact~\cite{bollen2009principal,kanellos2019impact}.
Consequently, a multitude of impact measures and
indicators have been introduced in the literature,
each of them better capturing a different aspect. 
In this work, we present an open dataset of different impact measures, which we freely provide.
We focus on measures that quantify three aspects of scientific impact, which can be useful for various real-life applications: \emph{popularity}, which reflects a publication's current attention (and its expected impact in the near future), \emph{influence} which reflects its overall, long-term importance, and \emph{impulse}, which better captures its initial impact during its ``incubation phase'', i.e., during the first years following its publication.

We focus on the first two aspects because they are well-studied: a recent experimental study~\cite{kanellos2019impact} has revealed the strengths and weaknesses of various measures in terms of quantifying popularity and influence.
Based on this analysis, and on a set of subsequent experiments~\cite{kanellos2020ranking}, we select the following measures: Citation Count, Page\-Rank, RAM, and AttRank. 
We also include the impulse, since it is a distinct impact aspect that is reflected by a Citation Count variant, which is utilised for the production of some useful metrics, like the FWCI\footnote{More info on this measure can be found in the Snowball metrics cookbook: \url{https://snowballmetrics.com}.}. This variant considers only citations received during the ``incubation'' phase of a publication, i.e., during the first $y$ years after its publication (usually, $y=3$); we refer to this useful Citation Count variant as \emph{Incubation Citation Count}, and we select it for inclusion in the BIP! DB collection, as well. 
In the next paragraphs, we elaborate on all the impact measures of our dataset.

\noindent \textbf{Citation Count (CC).}
This is the most widely used scientific impact indicator, which sums all citations received by each article. The citation count of a publication $i$ corresponds to the in-degree of the corresponding node in the underlying citation network: $s_i = \sum_{j} A_{i,j}$, where $A$ is the adjacency matrix of the network (i.e., $A_{i,j}=1$ when paper $j$ cites paper $i$, while $A_{i,j}=0$ otherwise). Citation count can be viewed as a measure of a publication's overall impact, since it conveys the number of other works that directly drew on it.

\noindent\textbf{``Incubation'' Citation Count (iCC).}
This measure is essentially a time-restricted
version of the citation count, where the time window
is distinct for each paper, i.e., only citations $y$ years
after its publication are counted (usually, $y=3$). The ``incubation'' citation
count of a paper $i$ is calculated as: $s_i = \sum_{j,t_j \leq t_i+3} A_{i,j}$, where $A$ is the adjacency
matrix and $t_j, t_i$ are the citing and cited paper's 
publication years, respectively.
iCC can be seen as an indicator of a paper's initial momentum (impulse) directly after its publication.

\noindent \textbf{PageRank (PR).} Originally developed to rank 
Web pages~\cite{page1999pagerank}, Page\-Rank 
has been also widely used to rank publications in citation
networks (e.g.,~\cite{chen2007finding,ma2008bringing,bip}). In this latter context, a
publication's PageRank score also serves as a measure of 
its influence. In particular, the PageRank score of a
publication is calculated as its probability of
being read by a researcher that either randomly selects
publications to read or selects publications based 
on the references of her latest read. Formally, the score 
of a publication $i$ is given by: 
\begin{equation}
    s_i = \alpha \cdot \sum_{j} P_{i,j} \cdot s_j + (1-\alpha) \cdot \frac{1}{N}
    \label{formula:pagerank}
\end{equation}
\noindent where $P$ is the stochastic transition
matrix, which corresponds to the column normalised version
of adjacency matrix $A$, $\alpha \in [0,1]$, and
$N$ is the number of publications in the citation network. The first addend of Equation~\ref{formula:pagerank} corresponds to the selection (with probability $\alpha$) of following a reference,
while the second one to the selection of randomly choosing any publication 
in the network. 
It should be noted that the score of each publication relies of the score of publications citing it (the algorithm is executed iteratively until all scores converge). 
As a result, PageRank differentiates citations based on the importance of citing articles, thus alleviating the corresponding issue of the Citation Count.

\noindent \textbf{RAM.} 
RAM~\cite{ghosh2011time} is essentially a modified Citation Count, where recent citations are considered of higher importance compared to older ones. Hence, it better captures the popularity of publications. This ``time-awareness'' of citations alleviates the bias of methods like Citation Count and PageRank against recently
published articles, which have not had ``enough'' time
to gather as many citations.
The RAM score of each paper $i$
is calculated as follows:

\begin{equation}
s_i = \sum_j{R_{i,j}}
    \label{formula:ram}
\end{equation}

\noindent where $R$ is the so-called Retained Adjacency Matrix (RAM) and $R_{i,j}=\gamma^{t_c-t_j}$ when publication $j$ cites publication $i$, and $R_{i,j}=0$ otherwise. Parameter $\gamma \in (0,1)$, $t_c$ corresponds to the current year and $t_j$ corresponds to the publication
year of citing article $j$.

\noindent \textbf{AttRank.} 
AttRank~\cite{kanellos2020ranking} is a Page\-Rank variant that alleviates its bias against recent publications (i.e., it is tailored to capture popularity). AttRank achieves
this by modifying PageRank's probability of randomly 
selecting a publication. Instead of using a uniform probability,
AttRank defines it based on a combination of the publication's 
age and the citations it received in recent years. 
The AttRank score of each publication $i$ is calculated based on:

\begin{equation}
s_i = \alpha \cdot \sum_{j} P_{i,j} \cdot s_j
    + \beta \cdot Att(i)+ \gamma \cdot c \cdot e^{-\rho \cdot (t_c-t_i)}
    \label{formula:attrank}
\end{equation}

\noindent where $\alpha + \beta + \gamma =1$ and 
$\alpha,\beta,\gamma \in [0,1]$. $Att(i)$ denotes 
a recent attention-based score for publication $i$, which
reflects its share of citations in the $y$ most recent
years, $t_i$ is the publication year of article
$i$, $t_c$ denotes the current year, and
$c$ is a normalisation constant. Finally, 
$P$ is the stochastic transition matrix.

\section{The BIP! DB Dataset}
\label{sec:workflow}

\subsection{Data Collection \& Integration}
\label{sec:dataset}

BIP! DB's impact measures all rely on citation network analysis.
Hence, a major challenge in our work was to construct an interdisciplinary and as inclusive  as possible citation network on which all impact measures would be calculated. 
To achieve this, we gathered citation data and metadata from three data sources: OpenCitations'~\cite{opencitations} COCI dataset, Microsoft's Academic Graph (MAG)~\cite{Sinha2015,Wang2020}, and Crossref~\cite{Hendricks2020}. The current dataset version (regular updates are scheduled in the future) exploits the latest version of the COCI dataset (Sep 2020), and recent snapshots of MAG (Aug 2020) and Crossref (May 2020). Our dataset production workflow collects, cleans, and integrates data from these sources to produce a citation graph based on the distinct DOI-to-DOI relationships found. Since the publication year is required for some of the measures to be calculated, publications lacking this information were excluded from the final network. Table~\ref{table:statistics} summarises some statistics for the original data sources and the complete, integrated dataset.

\subsection{Calculation of Impact Measures}

As discussed in Section~\ref{sec:dataset}, the volume 
of processed data exceeds $100$ million publications and 
$1$ billion references.
Hence, particular care must 
be taken by any algorithms developed for the calculation 
of the required impact scores, to allow for the handling 
of time-efficient and scalable updates.

By examining the formulas of the impact measures presented 
in Section~\ref{sec:bgr}, we can observe that all of them
rely on the analysis of the underlying citation 
network, by calculating a (possibly weighted) sum of 
scores, which are received by each publication by its citing
articles. 
Hence, we can take advantage of data parallelism when implementing the algorithms calculating these measures.
In particular, each measure can be implemented as a set of MapReduce 
operations where each publication \emph{maps} its score (e.g., 
a citation, its PageRank score, etc) to its cited
papers. The final score of each publication, in turn,
results from an aggregation (\emph{reduce}) of all 
scores mapped to it. 
Additionally, PageRank, and AttRank in
particular, are iterative processes, which require such 
sets of operations to repeat until the calculated publication
scores converge. Therefore, we chose Spark, which is 
particularly suitable for data parallel iterative processes, as our development platform of reference.
In particular, we implemented all algorithms as PySpark scripts, running on Spark version 2.3.
All impact measure calculations are performed on a cluster of $10$~VMs, each with $4$~cores and $8$GB RAM, and each script runs with up to 35 Spark workers.

\subsection{Published Data Records}

\begin{table}[!t]
    \centering
    \begin{tabular}{l c c}
    \textbf{Dataset} & \textbf{DOIs} &\textbf{Citations} \\ \hline \hline
    COCI & $59,455,882$ & $ 733,366,727 $
     \\ \hline
    CrossRef  & $96,703,144$ & $596,803,579$
     \\ \hline
    MAG &  $90,224,789 $ & $ 1,177,733,277$
     \\ \hline
    Unified Graph & $104,769,307$ & $1,254,817,030$
     \\ \hline
   \end{tabular}
    \caption{Distinct DOIs and citations per data source.}
    \label{table:statistics}
\end{table}

The current version of the BIP! DB dataset consists 
of five compressed TSV files, one
for each impact measure provided. All files follow the same
format: each line contains two data columns, where the 
first corresponds to the DOI of a publication, followed by
the column which corresponds to the score of the measure.
    
For the sake of clarity, each of the files published contains in its name the algorithm configuration that
produced the respective measure scores (the parameters were selected according to previous experiments, e.g.,~\cite{kanellos2019impact}). 
For example, the file named ``PR\_graph\_uni\-verse2\_1.txt\_a0.5\_error1e-12.gz''
contains the Page\-Rank scores calculated with parameter
$\alpha=0.5$ and with a convergence error set to 
$\epsilon \leq 10^{-12}$. All files published are 
freely available in Zenodo\footnote{BIP! DB dump, \url{https://doi.org/10.5281/zenodo.4386934}} under the Creative Commons Attribution 4.0 International license.

\subsection{Updated BIP! API}

Extra effort was given to update the existing BIP! API~\cite{bip} with the most recent version of the BIP! DB dataset so to provide programmatic access to the impact measures of the same set of publications. 
As a result, all calculated impact measures are also accessible via a public REST API\footnote{BIP! API documentation, \url{https://bip-api.imsi.athenarc.gr/documentation}}. 
It supports retrieving impact scores for a given article or for a number of articles given a list of DOI identifiers. 
The API response includes all five impact scores in a simple JSON object, 
one for each requested DOI. 

\section{Discussion}
\label{sec:disc}

To highlight the fact that the different
measures capture semantically diverse impact aspects we present, in Table~\ref{table:correlations}, the pairwise
top-$k$ correlations\footnote{We use Spearman's
top-$k$ rank correlation $\rho_{min}$ as defined in ~\cite{fagin2003comparing} ($\rho \in [-1,1]$,
with $-1, 0, 1$ indicating 
perfect inverse, no, and perfect correlation, respectively).} of the top-ranking $1,000,000$ papers 
(corresponding roughly to the top-$1\%$ of papers)
for each pair of impact measures we calculated.
Intuitively, we expect to see high correlations between measures that capture the same impact aspect. In general, our findings confirm this intuition since the popularity measures (AttRank \& RAM) appear to be highly correlated ($\rho>0.9$), while the influence ones (CC \& PR) appear to have a moderate correlation ($\rho>0.4$). 
Of course, as discussed, there are differences between measures of the same aspect, thus we 
do not expect pairs of measures for the same impact aspect
to correlate perfectly. In addition, in this experiment we only examine the correlation of the top-$1\%$ publications; the full set would reveal larger correlations for the measures of the same aspect (e.g., using the top $10\%$ publications we measured $\rho$ values greater than $0.6$ for CC and PR). 
Finally, based on the same measurements, iCC at best correlates weakly ($\rho<0.4$) to other measures, supporting the intuition that it captures a distinct impact aspect (impulse).

\begin{table}[!t]
    \centering
    \begin{tabular}{c | c c c c c}
    \textbf{} & \textbf{iCC} & \textbf{CC} & \textbf{PR} & \textbf{AttRank} & \textbf{RAM} \\ \hline \hline
    \textbf{iCC} & $1$ & $0.0985$ & $-0.3468$ &	$0.3141$ & $0.3042$ \\
   \textbf{CC} & & $1$ & $0.4144$ & $0.4583$ & $0.2774$
   \\
   \textbf{PR} & &  & $1$ & $ -0.0675 $ & $-0.2598$
   \\   
  \textbf{AttRank} & &  &  & $1$ & $0.9056$ \\
  \textbf{RAM} & &  &  &  & $1$ \\  
   \end{tabular}
    \caption{Top-$1\%$ pairwise correlations of impact measures.}
    \label{table:correlations}
\end{table}

Our data are openly available both on Zenodo
and through an open API, so to facilitate their utilization by third-party research teams and to enable building useful 
services on top of them. 
Furthermore, the files in the repository are available in TSV format, to allow for easy editing and/or processing for import in various database management systems. 
Finally, in line with the need for any relevant applications to use fresh data, we plan to update the provided files regularly, taking into consideration the update rate of the integrated data sources. 

There are many possible applications that can leverage the data provided by BIP! DB. For instance, academic search engines may utilise impact measures to rank publications based on different impact aspects\footnote{Our academic search engine BIP! Finder~\cite{bip},
is an 
attempt in this direction.}, while science monitors (like the Open Science Observatory~\cite{os_obs}) may use them to produce intuitive reports and visualisations. Furthermore, publication-level impact measures can be propagated and aggregated to related entities (e.g., datasets, software packages, individual researchers) to quantify their expected impact, which may be useful for various applications (e.g., planning for officers in funding organisations, decision support for HR departments of research institutions). 
Finally, the calculated impact measures may be used as features in various machine learning applications that apply data mining on citation networks. 

At this point, we would like to highlight that researchers should always have a large toolbox of impact measures available, in order to get the full picture about a publication's impact and to successfully secure themselves from various types of attacks and malpractices in the filed of research assessment. 
This is why we envisage to continuously update and extend the BIP! DB dataset to always contain up-to-date data and to be inline with the latest developments in the field of research assessment and scientometrics. 
To this end, we plan to update our set of included measures with new ones performing better in terms of effectiveness, and to extend BIP! DB to include measures that reflect additional impact aspects. 
Finally, since scientific impact is not always entirely correlated with 
publication quality, we intend to also include measures that capture other aspects of a publication's merit (e.g., novelty, readability). In general, contrary to other works, that construct and make available unified citation graphs (e.g.,~\cite{opencitations,dimensions}), the focus of our work is not on providing the unified graph itself, but on producing an open dataset of easily interpretable scientific impact measures, calculated on a unified graph, which can satisfy diverse needs and are ready to use.

\section{Conclusions}
\label{sec:concl}

We presented BIP! DB, an open dataset containing various impact measures, calculated for hundreds of millions of scientific articles. 
Our dataset provides a multidimensional view of article impact, and thus may be potentially beneficial for many different applications and diverse stakeholders.
Furthermore, we aim to deliver regular updates of our dataset in line with the updates of the data sources we use. Finally, in the future, we additionally plan to extend the published dataset not only with further impact measures, but also with other indicators capturing aspects other than the ones strictly related to scientific impact, such as readability and novelty.

\begin{acks}

This project has received funding from the European Union’s Horizon 2020 research and innovation programme under grant agreement No 101017452.

\begin{figure}[!h]
     \centering
     \includegraphics[width=0.1\linewidth]{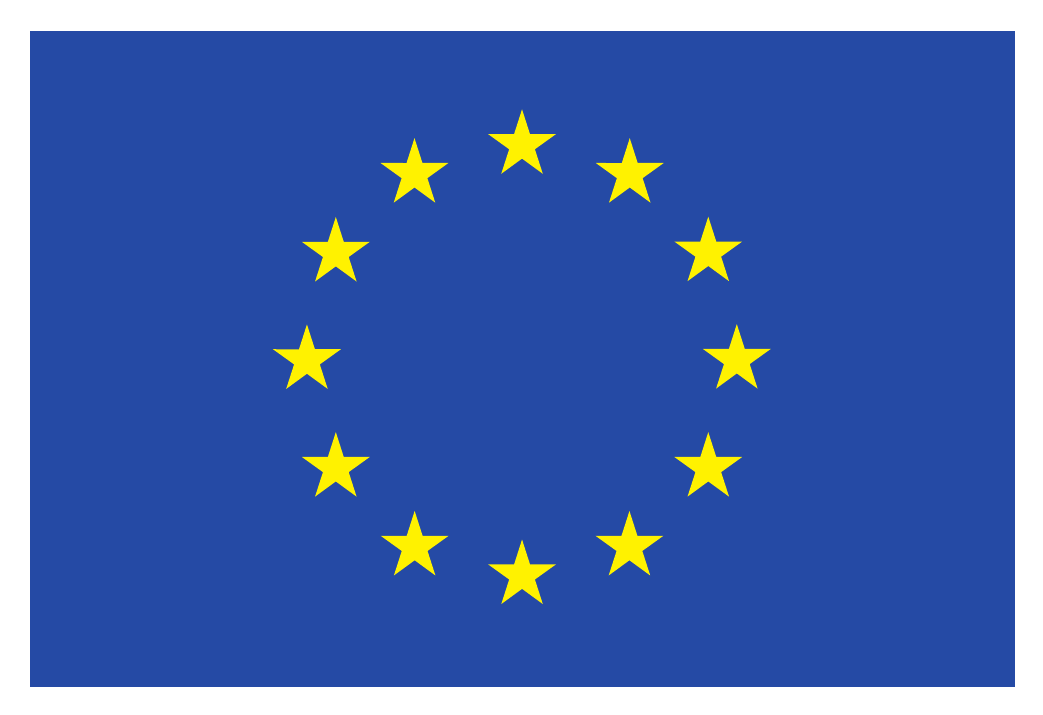}
\end{figure}

\end{acks}

\bibliographystyle{ACM-Reference-Format}
\bibliography{main}


%

\end{document}